RESEARCH ARTICLE

# Category learning can alter perception and its neural correlates

**Fernanda Pérez-Gay Juárez**[1,2]*, **Tomy Sicotte**[2], **Christian Thériault**[2], **Stevan Harnad**[1,2,3]

1 McGill University, Montréal, Canada, 2 Université du Québec à Montréal, Montréal, Canada, 3 University of Southampton, Southampton, United Kingdom

* fernandapgj@gmail.com, fernanda.perezgay@mcgill.ca

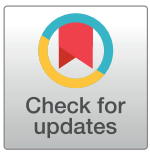

**Data Availability Statement:** All relevant data are within the manuscript and its Supporting Information files.

**Funding:** This work was supported by: SH - Fonds de recherche de Quebec Nature et Technologies; FPGJ - Fonds de recherche de Quebec Nature et TEchnologies - Merit scholarship for foreign students. The funders had no role in study design, data collection and analysis, decision to publish, or preparation of the manuscript.

## Abstract

Learned Categorical Perception (CP) occurs when the members of different categories come to look more dissimilar ("between-category separation") and/or members of the same category come to look more similar ("within-category compression") after a new category has been learned. To measure learned CP and its physiological correlates we compared dissimilarity judgments and Event Related Potentials (ERPs) before and after learning to sort multi-featured visual textures into two categories by trial and error with corrective feedback. With the same number of training trials and feedback, about half the subjects succeeded in learning the categories ("Learners": criterion 80% accuracy) and the rest did not ("Non-Learners"). At both lower and higher levels of difficulty, successful Learners showed significant between-category separation—and, to a lesser extent, within-category compression—in pairwise dissimilarity judgments after learning, compared to before; their late parietal ERP positivity (LPC, usually interpreted as decisional) also increased and their occipital N1 amplitude (usually interpreted as perceptual) decreased. LPC amplitude increased with response accuracy and N1 amplitude decreased with between-category separation for the Learners. Non-Learners showed no significant changes in dissimilarity judgments, LPC or N1, within or between categories. This is behavioral and physiological evidence that category learning can alter perception. We sketch a neural net model predictive of this effect.

## Introduction

The linguists Sapir (1929) and Whorf (1940; 1956) suggested that the language we speak shapes the way we see the world. According to this "linguistic relativity" hypothesis, it is learning to put things into different categories by giving them different names that makes them look more distinct to us, rather than vice versa [1]: for example, the rainbow looks to English-speakers as if it were composed of qualitatively distinct color bands because of the way English subdivides and names the visible wavelengths of light; the different shades of green all look like greens rather than blues because in English we learn to call them "green" rather than blue. In languages that use the same word for green and blue (the equivalent of "grue," [2]), the speakers would see only one qualitative "grue" band in the rainbow, instead of a green and a blue one.

It has turned out, however, that qualitative color categories are not perceptual effects induced by category naming. The anthropologists Berlin and Kay showed that basic color perception is universal, irrespective of the names and subdivisions assigned by different languages [3]. Visual neurophysiology has confirmed that the colors we see and the boundaries between them are determined by inborn neural feature-detectors: The cones in our retinas are selectively tuned to the red, green and blue regions of the frequency spectrum and our visual cortex has color-sensitive neurons responsible for paired red/green and blue/yellow opponent processes [4–6]. Hence the perceived qualitative differences among colors are not the result of language but an inborn consequence of Darwinian evolution. Is this enough to demonstrate that Whorf and Sapir were wrong about the effects of naming on perception? To answer this question we must consider an activity more basic than naming, and a prerequisite for it: categorization.

To *categorize* is to do "the right thing with the right kind of thing": responding to things differentially, manipulating them adaptively, sorting them into groups and giving them different names [7,8]. According to the "classical view" of categorization [9], what determines whether something is or is not a member of a category is the features that *covary* with membership in the category: present in members, absent in non-members. Features are initially sensory properties of things, such as size, color, shape, loudness or odor.

*Categorical perception* (CP) is a perceptual phenomenon in which the members of different categories are perceived as more dissimilar (between-category separation) and/or the members of the same category are perceived as more similar (within-category compression) than would be expected on the basis of their physical features alone [10–12]. The rainbow effect in color perception is actually a striking example of CP: The wave-length difference between a blue and a green looks much bigger than *an equal-sized wave-length difference* between two shades of blue within the blue band. Color CP, however, is, as noted above, dependent on inborn feature-detectors and hence not directly related to language or learning. To test the Whorf-Sapir hypothesis the right question to ask is: what happens with the categories that we have to learn through experience?

If we open a dictionary, we encounter mostly names of categories that we had to learn through either direct experience or verbal instruction [13,14]. It is very unlikely that we were born with innate detectors for all these categories. If, as suggested by the classical view of categorization, we need to detect the features that distinguish category members from nonmembers so that we can do the right thing with the right kind of thing, then with categories for which we have no inborn feature-detectors our brains need to *learn* to detect the features [15,16].

Many categories are obvious, or almost obvious: The differences between members and non-members already pop out. There is no need for learned CP separation/compression to distinguish zebras from giraffes: Their prominent natural difference in shape and color is enough. But the obvious similarities and differences in the sensory appearances of things are not always enough to guide us as to what to do with what—at least not for some categories, and not immediately. For categories whose covarying features are harder to detect (rather than evident upon repeated exposure without corrective feedback), learning to categorize may be more challenging and time-consuming.

Learned CP occurs when category learning induces between-category separation and/or within-category compression (the category boundary effect). This effect is not based on comparing perceived differences between and within categories for equal sized physical differences, as with colors and phonemes. It is based on comparing perceived differences between and within categories *before and after having learned the categories*.

It is important to distinguish CP induced by category learning from increased distinctiveness overall induced by mere repeated exposure, without corrective feedback. [17]. Classical acquired distinctiveness of *stimuli* (not categories), induced by repeated exposure [18] (unsupervised/unreinforced learning based on feature and feature-co-occurrence frequencies) makes all stimuli look more distinct from one another, like an expanding universe. Acquired distinctiveness of categories (comparing stimuli within and between categories) can arise through unsupervised learning if the categories are already well separated in sensory space by salient natural sensory discontinuities (like mountains vs plains vs valleys). But if the features distinguishing the categories are not already salient, trial-and-error with error-corrective feedback (supervised learning) is necessary to learn them. With acquired distinctiveness *between* categories (and acquired similarity *within* categories) the outcome is not the result of an expanding-universe effect in which all stimuli become more distinct from one another; instead, stimuli become more distinct from one another if they are in different categories (and sometimes they also become more similar to one another if they are in the same category).

Hence learning some categories does not generate category-specific CP, but merely an overall increase in all interstimulus distances [19–23] whereas learning other categories does generate CP [24–31]. Studies vary in the stimuli and tasks they use and how they measure CP effects (e.g., via similarity judgments, psychophysical discriminability or electrophysiological correlates); and CP effect-sizes vary considerably across studies in the relative degree of separation or compression they induce [32]. But the learned CP effect itself seems to be real. The question is: what factors induce it, and why?

Most authors attribute learned CP effects to feature-detection. A variety of psychophysical studies have shown that learning the features (or dimensions) relevant to category membership increases perceptual and attentional sensitivity to those features, resulting in easier detection [29,33]. The feature-detector may act as a filter, altering the perceived similarity between and within categories to make category members "pop out" [34–36] so that we can reliably go on to do the right thing with them.

Some learned CP effects are still open to the interpretation that they are not perceptual changes but a response bias from having learned to name the category ("naming bias" or "category label bias"): a tendency to judge things as less similar when their names are different and more similar when their names are the same [37–39]. One way to test whether CP effects are perceptual or verbal is to analyze brain activity during category learning.

Recent research in visual neuroscience suggests that early perceptual systems are not hard-wired; they can be tuned by several types of information, including attention, expectation, perceptual tasks, working memory and motor commands [40]. These modifiable properties become important in extracting relevant information from the environment.

Among the learned CP studies cited above, some were accompanied by neuroimaging or electrophysiological analyses that detected neural changes induced by training. In a series of experiments, Sigala and Logothetis [41] have found that neurons in the inferior temporal cortex of monkeys could be selectively tuned to dimensions diagnostic of category membership, with enhanced neural activity in response to features relevant for categorization. These findings have been extended through non-invasive neuroimaging studies in humans [22,42]. Folstein and his team trained human subjects to categorize a series of cars, counterbalancing the relevant dimension across subjects. Having learned the category, subjects performed a match-to-location task inside the fMRI scanner: They were presented with two successive stimuli and asked to indicate only whether they appeared in the same location. The researchers found changes in the activity of both the anterior fusiform gyrus and the extrastriate occipital cortex when the cars differed on the category-relevant dimension (i.e., they belonged to different categories rather than the same category). This suggests that learning a category enhances the detectability of distinguishing features not only in the temporal association cortex that is related to high level visual processing, but also in earlier stages of perception that take place in the extrastriate visual cortex [29,43].

Event Related Potentials (ERPs), with their precise temporal resolution, provide information about the time course of stimulus processing: For example, semantic and visual processes during categorization can be dissociated in their time course as well as their location [44]. In category learning, ERP changes can help distinguish perceptual effects (earlier ERP components) from post-perceptual cognitive effects (later ERP components). Previous studies have found ERP correlates of categorization and category learning. [45–47], but to our knowledge none of them have explored their link with behavioural measures of CP.

### The present study

To test whether category learning induces between-category separation and within-category compression (the signature of CP), subjects were trained by trial and error with corrective feedback (supervised/reinforcement learning) to sort unfamiliar visual stimuli (black and white textures) into two categories. To avoid having local, verbalizable features, the textures were designed to generate a holistic perceptual effect. Pairwise interstimulus dissimilarity judgments (between categories and within categories) as well as scalp-recorded ERPs elicited by the stimuli were compared before and after learning the categories. A neural net model for category learning and feature filtering [48] predicted that the category learning would generate between-category separation and within-category compression. This prediction was confirmed, both perceptually and physiologically, first in an exploratory study and then in an independent replication (Experiment 2). The behavioral findings have been reported in a previous paper [48]. The present paper reports the physiological correlates of these behavioral findings and identifies the ERP correlates of learned CP. An increase in the perceived difference between members of different categories, accompanied by an increase in negativity in a perceptual component of the ERP (N1), occur after categorization training, but only in the successful learners. Those who fail to learn the category show no change in their perception, and no change in their N1.

## Experiment 1

### Materials and methods

This project was approved by the Comité institutionnel d'éthique de la recherche avec des êtres humaines, Université du Québec à Montréal, approval number 803_e_2017.

**Subjects.** Forty-two right-handed subjects (20 female, 22 male) aged 18–35 years were recruited online through Kijiji and the McGill Classified Ads website. They were either native English-speakers or native French-speakers and free of significant neurological or psychiatric conditions. Each subject was assigned randomly to one of four levels of difficulty as described below. All subjects gave written consent.

**Stimulus generation.** To design a categorization task with unfamiliar stimuli and features that were distributed rather than local a large set of 270 x 270 pixel black and white square-shaped textures was generated (examples in Fig 1).

The building blocks of the textures were twelve 6 x 6 squares, each consisting of 18 black and 18 white pixels arranged in different patterns. These 12 squares were then paired arbitrarily, thus providing 6 pairs of mutually exclusive binary (0/1) micro-features (Fig 1). For simplicity, we will henceforth refer to these squares as "features". Each individual texture was thus built out of 900 features, 30 along the width dimension and 30 along the height dimension, their spatial positions distributed randomly. From left to right and

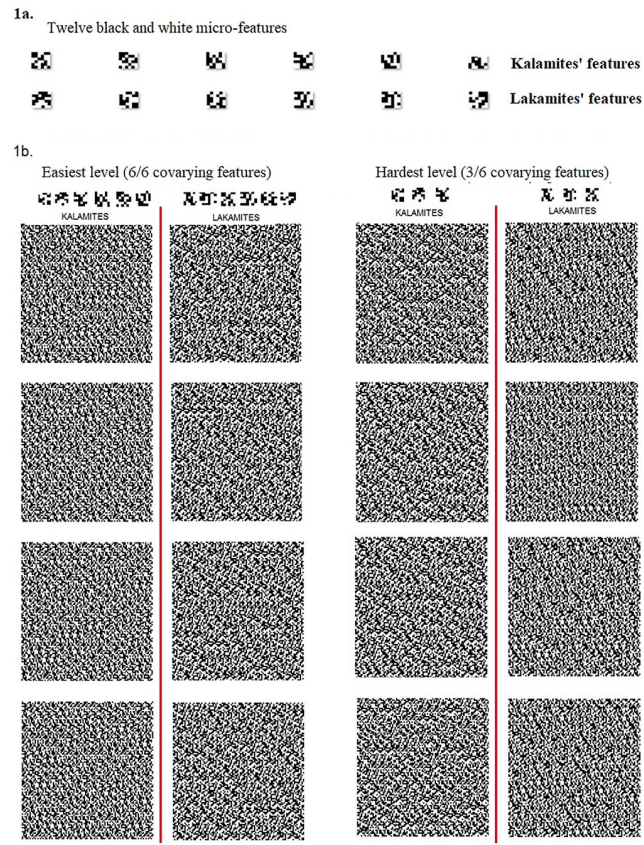

**Fig 1. Above (1a):** the six pairs of binary features used to generate the two texture categories: "Kalamites" (Ks) and "Lakamites" (Ls). **Below (1b)** Left: sample of 4 Kalamites and 4 Lakamites at the easiest level (6/6, in which all six features covaried with category membership) Right: 4 Kalamites and 4 Lakamites at the hardest level (3/6, in which only three of the six features covaried with membership; the non-covarying pairs varied randomly).

https://doi.org/10.1371/journal.pone.0226000.g001

top to bottom, a feature was added at random with replacement from the set of 6 features (all of the 6 features were equally represented in each stimulus) until an image of 17x17 features (510x510 pixels) was generated. The resulting 180x180 grid was then resized to 275x275 pixels using PIL. Image (ANTIALIAS) in Python 2.7, a high-quality filter based on convolution.

The stimuli were designed to produce four "a-priori" levels of difficulty. At the easiest level, all 6 binary features covaried with category membership: the 0-value of each binary pair occurred in every member of the K category (KALAMITES) and the 1-value of each pair occurred in every member of the L category (LAKAMITES). Our a-priori assumption was that stimuli in which all the features covaried with category membership would be the easiest to learn to categorize and that difficulty would increase as the proportion of covarying (relevant) features decreased and the proportion of non-covarying (irrelevant) features increased. The four levels of difficulty tested ranged from 6/6 co-variants (easiest), to 5/6, 4/6 and 3/6 (hardest). The non-covarying features varied randomly at each level, independent of category membership. Each set consisted of 180 different texture images, each of them presented two to three times for a total of 400 trials. Stimuli were generated using the PsychoPy2 open source software [49]. Although the proportion of covariant features (k/6) decreased at each difficulty level, only one arbitrary combination of k features was tested at each level, not every possible combination of k features: For example, all subjects trained at level 3/6 viewed stimuli with the very same three (arbitrarily chosen) covariant features (Fig 1).

**Procedure.** The experiment was conducted in a sound isolated chamber with dim lighting and no other sources of electromagnetic interference. Subjects were seated in a comfortable armchair in front of a glass window through which they saw the computer screen presenting the task. They had a keyboard placed on a table between them and the window to click on the K and L keys. Sixty-four electrode channels were used to record whole-head EEG data through the Biosemi Actiview2 amplifier. The task was built and presented using the PsychoPy2 psychology open source software [49].

**Task.** In this first experiment, the standard reinforcement learning task consisted of trial and error with corrective feedback. The training session lasted about forty minutes (pauses included). Subjects had to learn to categorize each texture as either a "KALAMITE" or a "LAKAMITE". Each set included one-hundred and eighty textures generated as described above. Subjects saw a total of four hundred textures (each stimulus appearing 2–3 different times during the task).

Each trial consisted of a fixation cross (500 ms) followed by one of the stimuli, shown at the center of the screen against a white background (1.25 s). Subjects were instructed to click K or L to indicate the category. They had to respond within 2s of the onset of the stimulus; if they did not, the computer prompted them to respond faster. Responses were followed by immediate feedback (lasting 750 ms) indicating whether the response had been correct or incorrect. Inter-trial interval was 2500 ms.

The 400 trials were divided into four blocks of a hundred stimuli each. Following each block, there was a pause in which subjects had to fill out a questionnaire asking whether they thought they had detected the difference between the KALAMITES and LAKAMITES. If they replied "yes", they were asked to describe what they were doing to categorize the stimuli. If they replied "no", they were asked to describe the provisional strategy they were using to try to sort them. Instructions and questionnaires were in English or French depending on the subjects' native language. Both responses and reaction times were reported during the task.

**Learning assessment.** The learning curves for all subjects were analyzed o determine which subjects had learned and which had not. The percentage correct in each series of 20-trial runs was calculated. Our criterion for successful learning was to reach and maintain 80% correct till the end of the 400 training trials, starting from at least 60 trials before the end. The point at which they reached the criterion was treated as as the "learning point" (see Fig 2, left). The subjects were accordingly divided into "Successful Learners" and "Non-Learners"

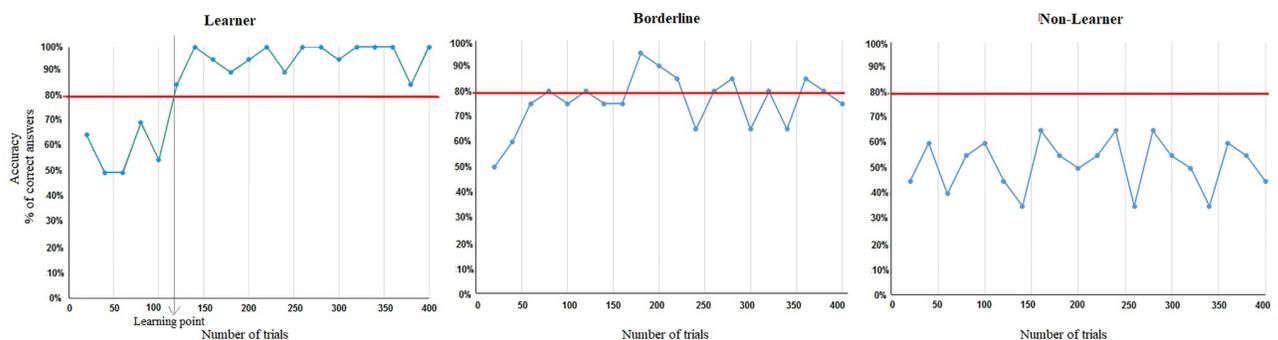

**Fig 2. Examples of learning curves for each of the three observed learning patterns.** From left to right: (a) Successful Learner, (b) "Borderline" and (c) Non-Learner. The red line corresponds to 80% correct. For the successful subjects, the point where they reach the 80% red line (if they stay above it thenceforward) is considered the "learning point," which then serves as a basis for splitting our EEG data for the before-after comparison.

https://doi.org/10.1371/journal.pone.0226000.g002

(right). However, at the higher difficulty levels, some subjects showed an unexpected learning pattern (middle): they reached 80% but then fell below it and kept rising above and below 80% till the end. These subjects were classified as "Borderlines" because they did not show a "Non-Learner" pattern (percent correct remaining around chance, 50%), but they didn't maintain our 80% criterion either.

To estimate degree of difficulty, we analyzed the number of trials required to reach the criterion and the percentage of Learners and Non-Learners for each set, assuming that with greater difficulty it would take more trials to reach criterion and fewer subjects would succeed in reaching and sustaining it.

**EEG Acquisition.** A Biosemi 64-electrode international reference cap was placed on the Ss' heads according to head circumference; electrodes were connected to the cap using a column of Conductive Gel to fill the gap between the skin and the electrodes. Six facial electrodes were placed at the common reference sites: two earlobes, above and below the right eye to record the VEOG (Vertical Electrooculogram), directly to the side of the left eye and directly to the side of the right eye to record the HEOG (Horizontal Electrooculogram). The signals were received by a Biosemi ActiveTwo amplifier at a sampling rate of 2048 Hz with a band pass of 0.01–70 Hz. Impedance of all electrodes was kept below 5kOhms. Data collection was time-locked to time point zero at the onset of visual stimulus presentation.

**EEG data analysis.** EEGLab 13.4.4b open source software [50] was used to process raw EEG files via the following steps: (a) The data were down-sampled to 500 Hz to decrease computational requirements. (b) A low pass (100 Hz) filter, high pass (3 Hz) filter and notch filter (60 Hz) were then applied. (c) Bad channels were identified by EEGLAB and were then interpolated. (d) The electrodes were re-referenced to a virtual average reference including all head electrodes but excluding the facial ones. (e) The data were divided into 3000 ms segments with individual epochs spanning from 1000 to 2000 ms around time zero. (f) A baseline was corrected based on the 200 ms before each stimulus onset. (g) EEGLAB function Runica [51] was used to identify independent components. (h) The first 10 components were visually inspected. Components associated with blinks and eye-movements according to topography and power spectrum were rejected. (i) The data were then separated into two parts—before learning and after learning for the Learners or first-half and second-half for other subjects. (j) Noisy epochs were filtered using an extreme value filter (+/- 100 μV) and then a probability filter with a 2 standard deviation limit for single channel and a 6 standard deviation limit across channels.

For the ERP analysis, Successful Learners' data was divided into two segments, based on the point when the subject reached (and sustained) our 80% learning criterion, as illustrated in Fig 2. The average ERP waveform elicited by the stimuli for the trials before and after this point was compared. For Non-Learners the datasets were divided in half and the first half of the trials was compared to the second half to control for ERP effects that were not due to learning to categorize (i. e. mere exposure/repetition effects). Once the datasets were split, grand averages for comparisons within subjects (before vs. after learning or first half vs. last half trials) and between subjects (Learners vs. Non-Learners) were computed. Limitations of this approach are considered in the Discussion section. (Reported in S1 File is an alternative approach in which subdivided both Succesful Learners' and Non-Learners' trials block by block).

ERPs from -200 to 1100 ms were plotted around time zero. Statistical analyses were conducted using the EEGLab software (parametric statistics, p<0.05, with Bonferroni correction). After identifying our Regions of Interest and significant time windows (see below), mean ERP voltages were measured in time windows centered on the peak of each component of interest. Amplitude (mean voltage) differences within subjects were assessed with student t distributions; effect sizes were calculated using Cohen's d and differences between subjects were assessed with repeated-measures ANOVA, all using the IBM SPSS 23 Statistical Software. Scalp distributions were plotted for each condition in the time-windows of interest.

## Results

**Analysis of learning.** Forty-two subjects (aged 19–34, 22 male, 20 female) completed the visual category-learning task; each was assigned to one of our four difficulty levels (Table 1). Overall, 28 of the 42 subjects successfully attained our a-priori criterion (reaching and maintaining at least 80% correct for at least the last 60 trials); four additional subjects were classed as Borderlines. The remaining eleven subjects did not reach the learning criterion throughout the task and were classed as Non-Learners.

The number of trials it took to learn to categorize as well as the overall accuracy through the categorization task for each difficulty level (Table 2) were examined. A one-way ANOVA with difficulty as between-subjects factor revealed that the number of trials it took to learn did not differ significantly between difficulties ($F(3,40) = 0.840$, $p = 0.481$; linear contrast $F(1,40) = 0.006$, $p = 0.940$), while the mean accuracy did ($F(3,40) = 5.576$, $p = 0.003$; linear contrast, $F(1,40) = 0.072$, $p = 0.789$). A HSD-Tukey post-hoc analysis of the accuracy between difficulties revealed the only significant difference was between level 2 (5/6) and level 3 (4/6): mean difference = 16.09%, $p = 0.013$. A detailed analysis of the difficulty assessment has already been reported in a previous paper (Pérez-Gay, et al., 2017).

Repeated-measures ANOVAs tested how Reaction Times and Response Accuracy changed across the four successive blocks. An interaction between block and Learning group showed that reaction times and accuracy across blocks were significantly different between Learners, Non-Learners and Borderlines (Accuracy: Wilks' Lambda = 0.560, $F(6,72) = 4.041$, $p = 0.02$, $\eta2 = 0.252$; reaction times: Wilks' Lambda = 0.622, $F(6,74) = 3.212$, $p = 0.008$, $\eta2 = 0.221$).

For Learners, response accuracy increased linearly (Wilks' Lambda = 0.917, $F(3,24) = 88.420$, $p<0.01$, $\eta2 = 0.917$; linear contrast, $F(1,26) = 170.022$, $p<0.001$, $\eta2 = 0.867$) and reaction times decreased linearly (Wilks' Lambda = 0.359, $F(3,24) = 14.307$, $p<0.001$, $\eta2 = 0.641$; linear contrast, $F(1,26) = 29.166$, $p<0.001$, $\eta2 = 0.529$). This pattern was absent in the Non-Learners whose accuracy did not change significantly across blocks (Wilks' Lambda = 0.710, $F(3,8) = 1.091$, $p = 0.407$, $\eta2 = 0.290$), and whose Reaction Times changed, but not linearly (Wilks Lambda = 0.318, $F(3,8) = 5.714$, $p = 0.022$, $\eta2 = 0.682$,; linear contrast, $F(1,10) = 0.871$, $p = 0373$, $\eta2 = 0.080$).

**Table 1. Outcome profile for each a-priori difficulty level in Experiment 1.**

| A-priori difficulty | Covarying features | Learners | Borderline | Non-Learners | Trials to learn: mean (SE) | Accuracy: mean (SE) |
|---|---|---|---|---|---|---|
| 1 | 6/6 | 7 | 1 | 3 | 140 (20) | 79% (4.1) |
| 2 | 5/6 | 8 | 0 | 3 | 194 (41) | 71% (4.7) |
| 3 | 4/6 | 5 | 3 | 2 | 278 (61) | 65% (3.04) |
| 4 | 3/6 | 8 | 0 | 2 | 138 (20) | 82% (3.7) |
| Total | | 28 | 4 | 10 | 173 (18) | 74%(2.2) |

https://doi.org/10.1371/journal.pone.0226000.t001

**Table 2. Number of learners and number of trials before reaching the learning criterion for easy and hard level.**

| Level | Immediate Learners | Successful Learners | Borderlines | Non-Learners | Trials to learn: Mean (SE) | Mean accuracy (SE) |
|---|---|---|---|---|---|---|
| Easier (5/6) | 6 | 10 | 0 | 5 | 106 (33) | 62% (3.17) |
| Harder (4/6) | 0 | 8 | 2 | 10 | 262 (32) | 81% (3.09) |
| Total | 6 | 18 | 2 | 15 | 175 (29) | 73% (2.66) |

https://doi.org/10.1371/journal.pone.0226000.t002

**ERP results.** In this first experiment, our goal was to explore the changes in early and late ERP components throughout the category learning task. Grand average ERPs were computed, combining the data from the four difficulty levels so as to have enough Learners and Non-Learners for comparison. As explained in the methods section, for the within-subjects analysis the data of the Learners were divided into trials before and after learning. The threshold and statistical methods described in Section 2.1.7 resulted in the rejection of an average of about 9 epochs (6%) from the before-learning trials (range: 2 to 26 trials, 1–10%) and about 16 epochs, (7%) from the after-learning trials (range: 1 to 38 trials, 2 to 12%) per subject.

For the Non-Learners, the first and second half of the trials was compared to rule out effects of repeated exposure. An average of about 14 epochs (7%) of the first half of the trials (range: 6 to 19 trials, 3–10%) and about 12 epochs (6%) of the second half of the trials (range: 9 to 22, 4–6%) per subject were rejected. Four subjects (3 Learners and 1 Non-Learner) were also excluded for for artifacts in more than 20% of the trials per condition or overall noisy recordings.

The midline electrodes Fz, Cz, Pz and Oz were examined visually as a first approach to assessing changes in ERP components after training (Fig 3). These plots revealed significant effects in an occipital N1 component and in a parietal Late Positivity, two components that have been reported as involved in category learning [47]. Mean voltages were extracted for these components in windows previously described in the literature (150–220 ms for the N1 [47,52–54] and 600–800 ms for the LPC [47,55,56]). Scalp distributions for after-minus-before difference waves (Fig 4) were then plotted.

**Effect of Learning on the occipital N1 (first negative peak).** Mean voltages (amplitudes) in the N1 window were extracted for the before-learning (first half) and after-learning (second

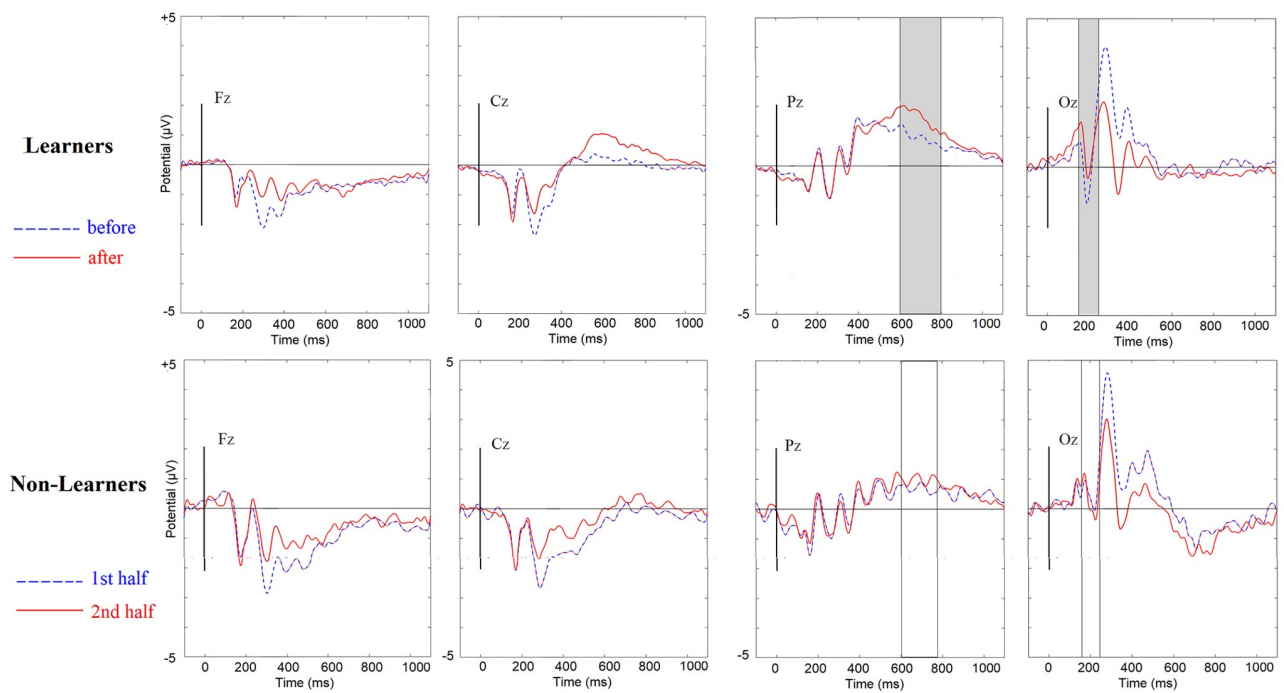

**Fig 3. Event Related Potentials in Mid-Line electrodes.** From left to right: Fz, Cz, Pz, and Oz. Positive is plotted up. A 30Hz low pass filter was used for plotting purposes only. **Upper:** Within-subject ERP grand averages for Learners (n = 23) across trials before vs. after reaching learning criterion. Highlighted in gray are time-windows with statistically significant differences. **Lower:** Within-subject ERP grand averages for Non-Learners (n = 9), for the first vs. second half of trials.

https://doi.org/10.1371/journal.pone.0226000.g003

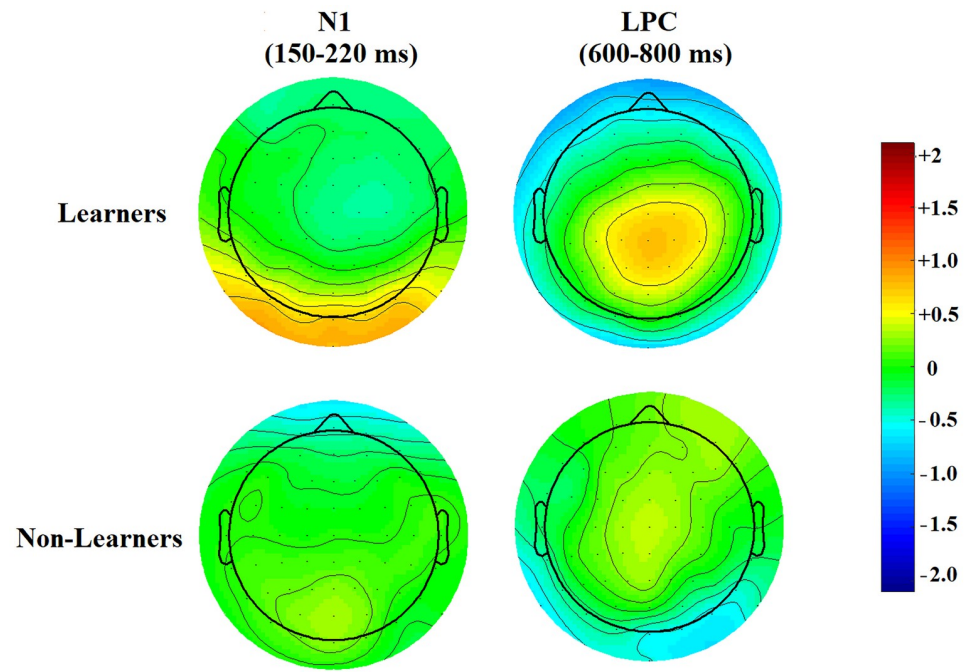

**Fig 4. Topological maps.** Topological heatmaps of after-minus-before difference waves in the N1 and LPC windows (Learners, upper row, Non-Learners, lower row). The vertical bar shows average voltage change.

https://doi.org/10.1371/journal.pone.0226000.g004

half) conditions in a cluster of five occipital electrodes (Iz, Oz, O1, O2. POz) and for each individual electrode in the cluster. A two-way mixed ANOVA with time (before reaching criterion vs. after reaching criterion [for Learners] or first half vs. second half [for Non-Learners]) as a within subject factor and group (Learners vs. Non-Learners) as a between-subject factor failed to show a significant interaction between learning to categorize and changes in the N1 amplitude in the occipital cluster ($F(1,32) = 1.605$, $p = 0.202$, $\eta2 = 0.026$).

Despite the absence of a significant interaction there were significant simple before-after effects within subjects for Learners only. These were noteworthy given the effect sizes. There was a statistically significant decrease in N1 negativity from before to after learning in our occipital cluster [mean change = 0.625, $t(24) = 3.406$, $p = 0.002$, Cohen's $d = 0.683$] for Learners but not for Non-Learners [mean change = 0.1516, $t(8) = 0.514$, $p = 0.621$, Cohen's $d = 0.171$]. (These effects were subsequently replicated independently and emerged statistically significant in Experiment 2).

**Effect of learning on the parietal LPC (late positive component).** The mean voltage (amplitude) in the LPC window before learning (first half) and after learning (second half) was extracted from a cluster of eight parietal electrodes (Pz, P1, P2, P3 P4, POz, PO3, pO4), and for each individual electrode in the cluster. A two-way mixed ANOVA with time (before/first half vs after/last half) as a within-subject factor and group (Learners vs. Non-Learners) as a between-subject factor showed a near significant interaction between learning to categorize and changes in the LPC amplitude in the parietal cluster ($F(1,32) = 3.867$, $p = 0.0058$, $\eta2 = 0.108$).

Post-hoc tests revealed a statistically significant increase in LPC positivity from before to after learning in the chosen cluster [mean change = 0.5325, $t(24) = 3.560$, $p = 0.002$, Cohen's $d = 0.712$] for Learners but not for Non-Learners [mean change = 0.0205, $t(8) = 0.514$, $p = 0.861$, Cohen's $d = 0.060$].

**Continuous correlational analysis.** To complement our discrete analysis (Learner vs. Non-Learner based on an 80% criterion) a continuous correlational analysis treated learning as a matter of degree instead of as all-or-none. Learners and Non-Learners were combined to measure the correlation between their performance (measured as percent correct in categorization on the last learning block) and the ERP changes (after-minus-before differences in N1 and LPC amplitudes).

Spearman's rank-order correlations between the size of the N1 change in the occipital cluster and accuracy in the last block of 100 trials were positive and significant: rho(34) = 0.359, p = 0.034, Fisher z = 0.3769, 95% CI = 0.024–0.622: The better the performance, the smaller the N1 amplitude after learning. The correlation between the change in LPC in the parietal cluster and the accuracy in the last block was also positive and significant (rho(34) = 0.366, p = 0.025, z = 0.3884, 95% CI = -0.012 0.578): The higher the performance, the bigger the LPC amplitude after learning.

## Experiment 2

Experiment 1 revealed two noteworthy ERP effects induced by learning to categorize novel visual textures: a decrease in negativity of an early (perceptual) component, N1, and an increase in positivity in a late (memory-related) component, the LPC. The data were analyzed in two complementary ways: (i) by partitioning our sample into two populations, Learners and Non-Learners, using a discrete performance threshold of 80% correct and (ii) by treating learning as continuous, combining all subjects and testing the correlation of each of the two ERP components of interest with their learning performance (percent correct) in the last learning block) with. To replicate and build on this initial outcome from exploratory Experiment 1, we did a second experiment using a new independent sample, testing the previous findings as a-priori predictions.

### Methods

**Subjects.** Forty-one right-handed subjects (23 Females, 18 Males) aged between 18 and 35 years were recruited online through Kijiji and the UQAM and McGill Classified Ads website.

**Stimuli.** For this experiment, feature-based a-priori levels were dropped because of feature inhomogeneity in favor of a-posteriori difficulty levels based on the mean accuracy throughout the task. The two stimulus sets that had turned out to be significantly different in terms of accuracy in Experiment 1 (Section 2.2.1) were used: one set that had proven easier, (5/6 covarying features) and one had proven harder (4/6 covarying features).

**Procedure.** The experiment was conducted in the same way as Experiment 1, including EEG recording. Dissimilarity judgements before and after learning from Perez-Gay et al (2017) [48] were added as described in the following section.

**Tasks.** In this second experiment, subjects made pairwise dissimilarity judgments on a subset of forty stimulus-pairs, once before the categorization training and once again after the training. They were presented with a fixation cross (500 ms), after which two stimuli appeared at the center of the screen for 1 s each, one after the other, with an inter-stimulus interval of 1 s. Subjects were then asked to rate their dissimilarity on a scale of 1 to 9, such that 1 corresponded to "very similar" and 9 to "very different". They were encouraged to make use of the full range of the scale. Of the total of forty pairs presented, 20 were within-category pairs (10 "Kalamites" and 10 "Lakamites") and 20 were between-category pairs (but of course before training subjects did not know the categories or their names). Responses and reaction times during the task were recorded. The same set of 40 stimulus pairs was presented in the same order for the dissimilarity judgements before and after training. Following the first set of dissimilarity judgements, subjects began their visual category training with corrective feedback, as in the first experiment (400 trials divided into four blocks with questionnaires in each pause).

**EEG Acquisition.** EEG data were collected and analyzed in the way described in Experiment 1.

**EEG data analysis.** EEGLab 13.4.4b open source software (Delorme and Makeig, 2004) was used to process the EEG files through the steps described in Experiment 1. Having identified our components of interest in the first experiment, the analyses focused on two main components. The first focus was on activity related to early visual processing (N1 component), for the region of interest: Electrodes O1, Oz, O2 and Iz analyzed in the time window between 150 and 220 ms. The other focus was on the Late Positive Component: electrodes CPz, Pz, P1, P2 and POz in the time window between 500 and 800 ms. The mean voltage in the time windows around the peak of each component of interest was extracted. Amplitude differences between conditions were assessed with repeated-measures ANOVA using the IBM SPSS 23 Statistical Software. scalp distributions for each condition in the time-windows of interest were also plotted.

**Analysis of dissimilarity judgements.** Average dissimilarity ratings for within-category and between-category pairs before ("pre") and after ("post") training were calculated for each subject, creating the variables "within-category pre-training" (Wpre), "within-category post-training" (Wpost), "between-category pre-training" (Bpre) and "between-category post-training" (Bpost). We created the variable difffW, for "within-category dissimilarity change" (Wpost–Wpre) and diffB for "between-category dissimilarity change" (Bpost–Bpre). Finally, a composite variable "Global CP" (diffB-diffW) was created. Its purpose was to serve as a single joint estimator of CP. Subtracting diffW from diffB will amplify Global CP if diffW and diffB go in opposite directions (within-category compression and between-category separation). If diffW and diffB have similar values, Global CP will be reduced.

The significance of these before/after changes was tested with repeated measures ANOVAs with "learning" as the between-group factor and t-tests to compare Successful Learners and Non-Learners separately. Pearson and Spearman correlations were calculated between the Learners' dissimilarity ratings after-minus-before learning (between categories as well as within) and the corresponding changes in their N1 and LPC.

## Results

**Analysis of learning.** Forty-one subjects completed the visual category-learning task, 21 assigned to the easier (5/6) and 20 to the harder (4/6) level. Twenty-four subjects successfully attained our learning criterion (reach and sustain 80% correct), 16 at the easier level and 8 at the harder level. A one-way ANOVA confirmed that the mean number of trials to criterion differed significantly between the levels ($F(1,16) = 11.034$, $p = 0.004$, $\eta2 = 0.408$), as did the Learners' overall accuracy ($F(1,16) = 13–242$, $p = 0.002$, $\eta2 = 0.453$) and accuracy in the fourth (last) block ($F(1,12) = 16.393$, $p = 0.001$, $\eta2 = 0.506$).

Interestingly, among the Learners at the easier level, seven already had an accuracy of over 80% from the very outset of the task. This suggests that, even without explicit instructions to categorize and without feedback, the mere exposure to the 40 pairs of stimuli during the preceding pairwise dissimilarity judgement task had been enough to induce passive learning in these subjects. These subjects were classed as Immediate Learners. Table 2 shows the outcome for the Easy (5/6) and Hard (4/6) level. At the hard level, two subjects were classed as "Borderlines." The remaining 15 subjects did not reach the learning criterion throughout the task and were classed as Non-Learners, five from the easier level and ten from the harder level (See Tables 1 and 2).

Repeated-measures ANOVAs tested the changes in Ss' Reaction Times and Response Accuracy through the four successive blocks. As in Experiment 1, there was a significant within-subjects effect of successive blocks on both learning measures for the Learners: response accuracy increased linearly across blocks (Wilks' Lambda = 0.199, $F(3,15) = 20.178$, $p<0.01$, $\eta2 = 0.801$; linear contrast, $F(1,17) = 60.232$, $p<0.001$) and reaction time decreased linearly (Wilks' Lambda = 0.288, $F(3,15) = 12.350$, $p<0.01$, $\eta2 = 0.712$; linear contrast, $F(3,15) = 18.98$, $p<0.001$). There was no effect for the Non-Learners (accuracy: Wilks' Lambda = 0.090, $F(3,12) = 83.774$, $p = 0.218$, $\eta2 = 0.165$; reaction times: Wilks' Lambda = 0.579, $F(3,12) = 2.907$, $p = 0.078$, $\eta2 = 0.421$).

A repeated measures ANOVA with difficulty as a between-subject factor for the Learners showed that there was no interaction between difficulty and block for reaction times ($F(1,16) = 0.701$, $p = 0.415$, $\eta2 = 0.042$) but there was a significant interaction for response accuracy ($F(1,16) = 13.731$, $p = 0.002$, $\eta2 = 0.462$), indicating that overall accuracy of the Learners was higher at the easier level.

**Dissimilarity judgements.** Subjects rated pairwise dissimilarity before and after the 400 training trials. If learning to categorize induces CP effects, we expect the dissimilarity score to increase for between-category pairs (between-category separation: positive diffB) and/or to decrease for within-category pairs (within category compression: negative diffW) for Learners only.

A three-way mixed ANOVA with time (pre-training vs. post-training) and pair type (between-category vs. within-category) as within-subject factors, group (Learner vs. Non-Learner) as a between-subject factor and dissimilarity judgement scores as the dependent variable was performed to assess the effects of training on perception. There was a statistically significant three-way interaction between time, group and pair type, $F(1, 31) = 0.273$, $p = .002$, partial $\eta^2 = 0.273$. There were three outliers, as assessed by inspection of boxplots. The outliers were kept in the analysis because they did not affect the results as assessed by a comparison of the results with and without the outliers.

Post-hoc tests revealed that Learners rated between-category pairs as significantly "more different" (positive diffB) after training compared to before (diffB) (between-category separation; mean diffB = 1.801, $t(17) = 6.453$, $p <0.01$, Cohen's d = 1.521). Learners also rated within-category pairs as less different after training compared to before (negative diffW), but this within-category compression was not significant (mean diffW = -0.43, $t(17) = -1.386$, $p = 0.184$. Cohen's d = 0.331). For Non-Learners there were no significant changes comparing before and after training, either between or within categories (mean diffB = 0.776, $t(14) = 1.935$, $p = 0.073$, Cohen's d = 0.504; mean diffW = 0.376, $t(14) = 1.609$, $p = 0.130$, Cohen's d = 0.244). Moreover, within-category changes (diffW) went in the opposite direction in Non-Learners, showing (non-significant) within category separation instead of compression. This suggests that mere exposure without category learning induces an overall expansion of the stimulus space in the Non-Learners, instead of the category boundary effects (separation/compression) observed in the Learners.

Independent samples t-test assessed differences in the composite "Global CP" variable described in the methods section. The results confirmed that the "Global CP" values were significantly higher for Learners (mean = 2.2347, SE = 0.4282) than for Non-Learners (mean = 0.400, SE = 0.262), $t(31) = 3.413$, $p = 0.038$.

According to a 2-phase neural network model for category learning [57] some learning can take place even during an "unsupervised" (autoencoding) phase of mere exposure, without any categorization training by trial and error with corrective feedback ("supervised learning"). We accordingly predicted and tested for a similar effect in our subjects. Fig 5 shows that the dissimilarity ratings for between-category and within-category pairs differed between Learners

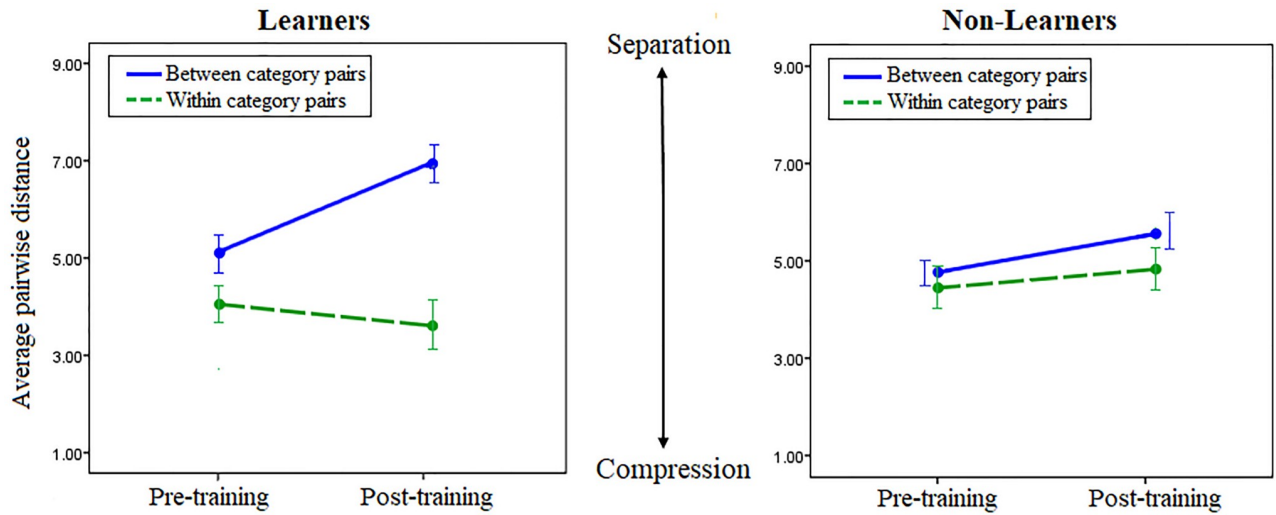

**Fig 5. Average perceived pairwise distance, between and within categories, before and after category learning, for Learners and Non-Learners.** Between-category (solid blue) and within-category (dashed green) dissimilarity ratings before and after training, for Learners (left) and Non-Learners (right), averaged across the two difficulty levels. Learners showed significant between-category separation. Non-Learners showed no significant changes. Error bars represent ±2 SE. (SE bar for between-category pairs of Non-Learners displaced to avoid overlap with within-category SE bar).

https://doi.org/10.1371/journal.pone.0226000.g005

and Non-Learners even before training. For Learners, between-category minus within-category dissimilarity before learning was significantly greater than for Non-Learners (mean between = 5.04, SE = 0.293; mean within = 4.01, SE = 0.294, t(17) = -4.645, p<0.001). For Non-Learners, there was no significant difference (mean between = 4.87, SE = 0.236 mean within = 4.55, SE = 0.231, t(13) = -1.482, p = 0.160). Like the pre-learning differences in the N1, these are evidence of unsupervised learning effects of mere passive exposure.

The differences between easier and harder conditions were tested with a four-way mixed ANOVA with time (pre-training vs. post-training) and pair type (between category vs. within category) as within-subject factors and group (Learner vs. Non-Learner) and difficulty (easy vs hard) as between-subject factors. There was a significant interaction of all these factors with the dissimilarity judgement scores (F(1,29) = 16.808, p<0.001, $\eta^2$ = 0.367). Post-hoc tests then tested the before-after differences for Learners between the easy and the hard level. While there was significant between-category separation in both the easier (mean diffB = 2.1625, t(9) = 5.955, p<0.001, Cohen's d = 2.714) and the harder condition (mean diffB = 1.35, t(7) = 3.359, p = 0.012, Cohen's d = 1.194), within-category compression was significant only in the easier condition (mean diffW = -1.145, t(9) = -3.359, p = 0.007, Cohen's d = -1.134). The harder condition showed only a small, non-significant separation for within-category pairs (mean diffW = 0.4562, t(7) = 1.165, p = 0.282, Cohen's d = 0.428). These results replicate the occurrence of between-category separation in both conditions.

As described in Experiment 1, an alternative way to analyze the effects of category learning on perception is to combine all subjects and treat the learning as a continuous matter of degree of accuracy (percent correct) for each subject instead of dividing the subjects into Learners and Non-Learners based on whether they met our (arbitrary) 80% criterion. Across all the subjects combined, all three of our CP measures—(1) the degree of after-minus-before separation between categories (positive diffB), (2) the degree of after-minus-before compression within categories (negative diffW), and (3) GlobalCP (DiffB minus Diff W)–were significantly correlated, in the predicted direction, with our measures of categorization accuracy (all four training blocks: diffB; rho = 0.422, p = 0.006, 95% CI = 0.116–0.674, diffW; rho = -0.442, p = 0.004,

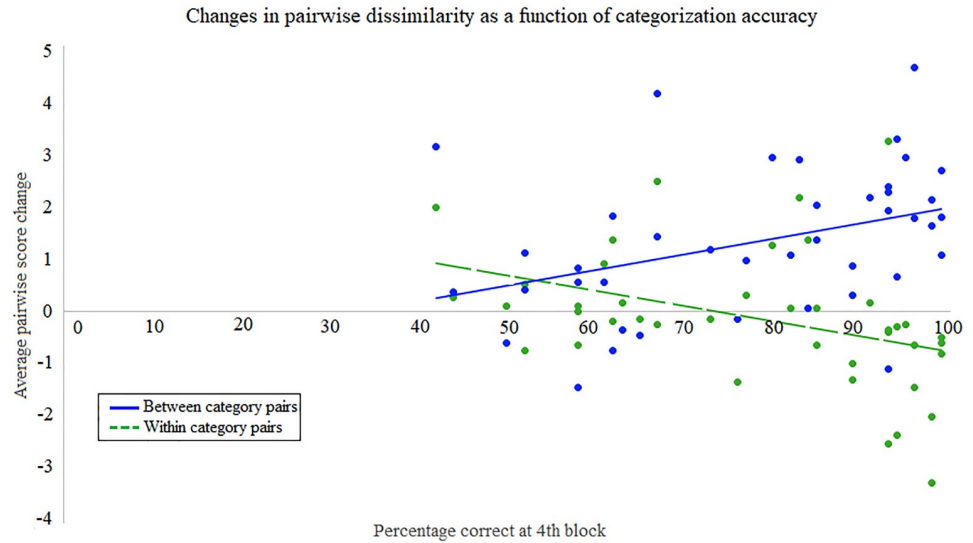

**Fig 6. Correlation between categorization accuracy in the last training block and changes in pairwise dissimilarity judgements (solid blue for between categories, dashed green for within categories) after-minus-before training.** The greater the categorization accuracy, the greater the increase in between-category category distance (greater perceived dissimilarity; separation) and the greater the decrease in within-category distance (lower perceived dissimilarity; compression).

https://doi.org/10.1371/journal.pone.0226000.g006

95% CI = -0.660 – -0.150, Global CP; rho = 0.630, p<0.001, CI: 0.370–0.803; just the last block: diffB; rho = 0.432, p = 0.005, 95% CI = 0.146–0.661, diffW; rho = -0.528, p<0.001, 95% CI = -0.715 – -0.265, GlobalCP; rho = 0.703, p<0.001, 95% CI = 0.465–0.841). Taken all together, these correlations provide robust evidence that the higher the categorization accuracy the greater the CP (Fig 6).

**ERP results.** For our ERP analysis, we focused on the two time-windows that showed significant effects associated with learning in the first experiment: The N1 and LPC windows. Two borderlines and the 5 Immediate Learners were excluded from the ERP analysis of the Learners. Grand averages were computed for the stimulus-locked ERPs, analyzing the within-subject changes before and after reaching the learning criterion. Using the threshold and statistical methods described in Section 2.1.7, an average of 10.67 epochs (5.56%) of the before-learning trials (range: 0 to 23 trials, 0–10.70%) and an average of 14.05 epochs, (5.99%) of the after-learning trials (range: 2 to 30 trials, 3.34 to 9.88%) per subject were rejected.

For Non-Learners, we analyzed the changes between the first and second half of the training trials. Using the threshold and statistical methods described in Section 2.1.7, an average of 13.36 epochs (6.71%) of the before-learning trials (range: 8 to 28 trials, 4.02–14.00%) and an average of 12.64 epochs, (6.30%) of the after-learning trials (range: 8 to 26 trials, 3.86 to 13.00%) per subject were rejected.

The mean voltage in a cluster of occipital electrodes (O1, Oz, Iz, O2) was extracted for N1 and in a cluster of parietal electrodes (Pz, P1, P2, CPz, CP1, CP2) for the LPC.

This experiment replicated the two main effects observed in Experiment 1: For the Learners, there was a decrease in the occipital N1(150–220) negativity and an increase in the LPC positivity from before learning to after. Both effects were absent in Non-Learners (Fig 7).

**Effect of Learning on N1:** Using the extracted mean voltage in the N1 window for the trials before learning (for the Learners; or the first half of all trials for the Non-Learners) and the trials after learning (second half of all trials for the Non-Learners) in a cluster of five occipital electrodes (Iz, Oz, O1, O2. POz), a two-way mixed ANOVA was performed with Time

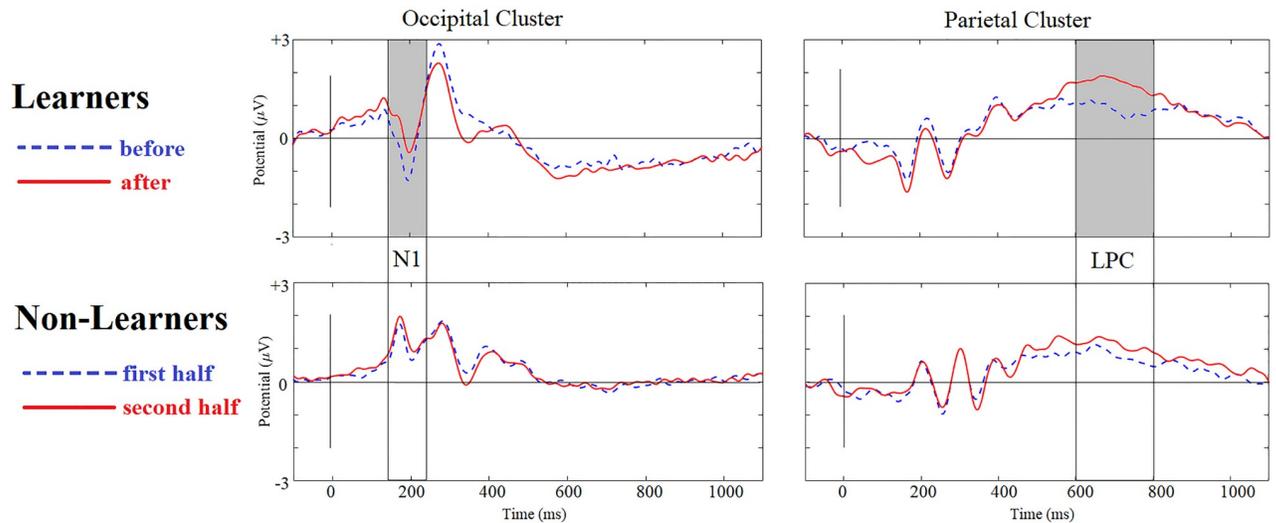

**Fig 7. N1 and LPC for Learners and Non-Learners.** Dashed blue: trials before learning for Learners (first half of trials for Non-Learners). Solid red: after learning for Learners (second half for Non-Learners). Positive is up. Grey indicates statistically significant regions within subjects in comparing before with after learning. There were three main findings: a decrease in N1 and an increase in LPC in Learners and the absence of any change in Non-Learners. A 30Hz lowpass filter was used for plotting purposes only.

https://doi.org/10.1371/journal.pone.0226000.g007

(before/first-half vs after/last-half) as a within-subjects factor and Group (Learners vs. Non-Learners) as a between-subject factor. There was a significant interaction between learning to categorize and changes in the N1 amplitude in this occipital cluster ($F(1,30) = 4.120$, $p = 0.0051$, $\eta2 = 0.121$).

Post-hoc tests confirmed that, for Learners, N1 amplitude decreased from before to after learning (mean amplitude change = 1.115, 0.002, Cohen's d = 1.004). Non-Learners showed no significant changes in N1 between the first and second half of the categorization trials (mean amplitude change = 0.095, $t(13) = 0.213$, $p = 0.802$, Cohen's d = 0.004). The continuous, correlational analysis combining Learners and Non-Learners showed a nearly significant, positive correlation between accuracy in the last block and changes in the N1 component, (rho = 0.327, $p = 0.061$, 95% CI = -0.117–0.648). The smaller the N1 amplitude (less negative), the bigger the accuracy in the last block.

As noted above, there was already a difference between the dissimilarity ratings of the Learners and the Non-Learners *before the categorization training*. A neural network model (to be described in more detail in the discussion section below) suggests an explanation for this unexpected finding: In the nets, which were, like the human subjects, trained to categorize by trial and error with corrective feedback, some learning can already take place during an unsupervised (autoencoding) phase, before the categorization training, through *mere repeated exposure, without categorization feedback*. Because of this pre-training effect, observed both for the nets and the experimental subjects in Experiment 2, absent in Experiment 1, the averaged ERP waveforms of the Learners during training before reaching our 80% criterion were compared with those of the Non-Learners. It turned out that the Learners' N1 amplitude on their training trials, before they had reached criterion, already differed significantly from the Non-Learners' N1 amplitude for the first half of their training trials, based on a between-groups one-way ANOVA (mean N1 amplitude difference = -3.9601, $F(1,31) = 7.454$, $p = 0.010$, $\eta2 = 0.199$). The only difference between Experiments 1 and 2 is that before their supervised learning (trial and error) phase the subjects in Experiment were exposed to 40 pairs of stimuli to be rated for Dissimilarity, whereas the subjects in Experiment 1 were not. The N1 difference

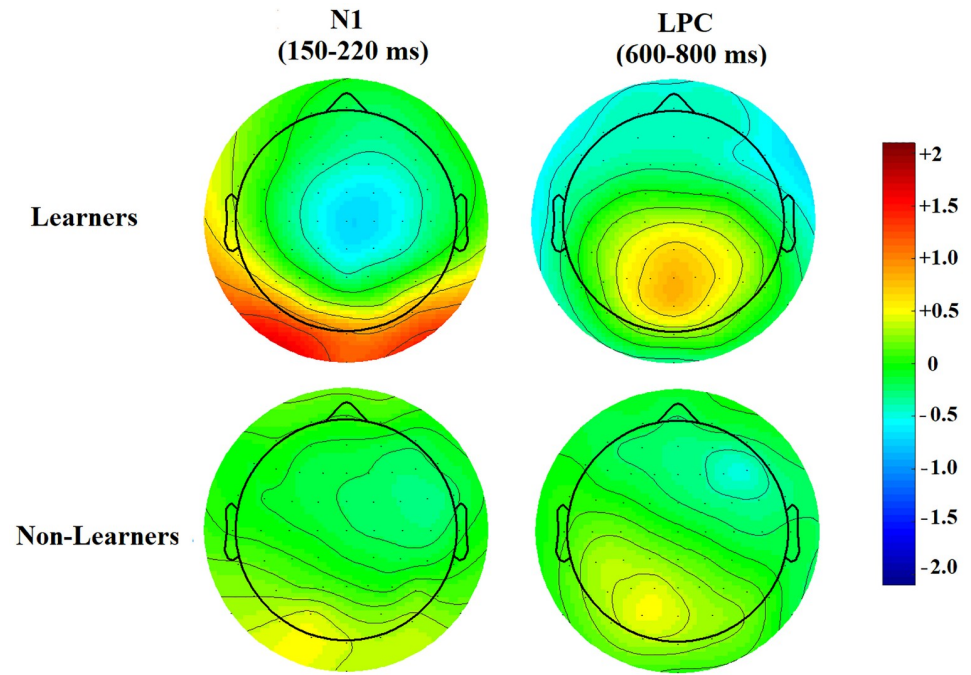

**Fig 8. Topological maps of after-minus-before difference waves in the N1 and LPC windows (Upper: Learners; Lower: Non-Learners).** Upper left shows the decrease in early negativity for Learners; occipital and parieto-occipital electrodes become significantly more positive after learning. Upper right shows an increase in late parietal positivity after reaching the learning criterion. None of these voltage changes are present in Non-Learners (lower row).

https://doi.org/10.1371/journal.pone.0226000.g008

observed (see Fig 7, first column, blue waveforms) suggests that *the Learners were already processing the features differently by the beginning of the training phase*. Moreover, in Experiment 2 there were 7 immediate Learners, performing above criterion from the outset of training. Our interpretation is that these subjects had already undergone some unsupervised learning during their passive exposure to the stimuli in the 40 dissimilarity-judgment trials before training.

**Effect of Learning on LPC.** A two-way mixed ANOVA with time (before/first-half vs after/last-half) as a within-subjects factor and group (Learners vs. Non-Learners) as a between-subjects factor used the mean voltage in the LPC window before learning (first half) and after learning (second half) for a cluster of eight parietal electrodes (Pz, P1, P2, P3 P4, POz, PO3, pO4). This analysis showed a significant interaction between learning to categorize and changes in the LPC amplitude in midline electrode Pz ($F(1,30) = 4.678$, $p = 0.039$, $\eta2 = 0.135$), although the effect did not reach significance in the entire parietal cluster ($F(1,30) = 1.708$, $p = 0.201$, $\eta2 = 0.054$).

Fig 8 shows the parietal topography of the LPC positivity increase and the occipito-parietal topography of the N1 negativity decrease for Learners.

The LPC effect was analyzed correlationally on all subjects combined, treating learning as a matter of degree (Fig 9). The continuous, correlational analysis of this late ERP effect in all subjects revealed that in the last training block parietal LPC amplitude correlated positively with percent correct (rho = 0.408, n = 33, $p = 0.018$, z = 0.4356, CI: 0.109–0.637) and negatively with average reaction time (rho = -0.495, n = 33, $p = 0.003$, CI: -0.717 –-0.144), replicating the outcome of Experiment 1. We infer that higher LPC amplitude reflects better learning (higher accuracy, shorter reaction times).

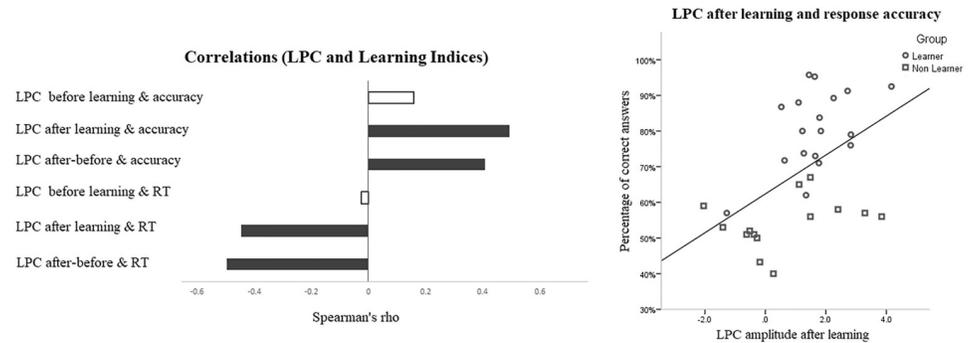

**Fig 9. Correlation of LPC with categorization accuracy (percent correct) and reaction time (RT).** Left: Spearman rank order correlations between LPC and accuracy and between LPC and RT for all subjects (Learners and Non-Learners) combined in the last training block: For Learners, "before" means the average before reaching the 80% success criterion, "after" means the average after reaching the criterion. For Non-Learners, "before" means the average for the first half of the training trials, "after" means the average for the second half. Dark bars are the significant correlations. Right: Scatter plot of the correlation between LPC and categorization accuracy on the last training block (Learners round, Non-Learners square).

https://doi.org/10.1371/journal.pone.0226000.g009

**ERPs and perceptual changes.** So far, we have reported ERP effects and perceptual changes induced by category learning separately. A series of Pearson product-moment correlations between the ERP parameters (N1 and LPC) and changes in perceived distance between and within categories was calculated to determine the relationship between the behavioral and the physiological variables.

**N1 amplitude after learning predicts between-category separation.** Tests for a relationship between changes in perceived distance within and between categories and N1 and LPC in Learners showed no significant correlation between either within-category compression (DiffW) or "Global CP" and N1 or LPC. Our main CP effect of interest, however, between-category separation (diffB), correlated positively, strongly and significantly with the amplitude of the occipital N1 after learning (rho = 0.603, n = 18, p = 0.003, 95% CI = 0.138–0.837): decreased N1 negativity predicted greater separation between categories in Learners. There was no correlation at all for Non-Learners.

The Learners' N1 correlations are shown in Fig 10. N1 after training (but not before) correlated significantly and positively with diffB (separation). N1 was uncorrelated with diffW (compression) both before and after training. LPC too was uncorrelated with separation or compression, before or after training. (Note that the p value for the correlation between N1 amplitude after learning and separation (diffB) is still significant after a Bonferroni correction for the 16 comparisons).

## Discussion

The objective of this series of experiments was to test whether CP could be induced by training subjects to sort unfamiliar visual stimuli into two categories through one hour of trial and error training with corrective feedback. We analyzed changes induced by category learning behaviorally (accuracy and dissimilarity judgments) and electrophysiologically (Event Related Potentials; ERPs) to test whether CP effects (between-category separation and/or within-category compression) are induced by category learning, whether they are perceptual, and what role they play in the mechanisms underlying category learning.

Our findings can be summarized as follows: Experiment 1 revealed electrophysiological correlates of category learning: a decrease in occipital N1 amplitude (an early perceptual component) and an increase in parietal LPC amplitude (a later decisional component) in those

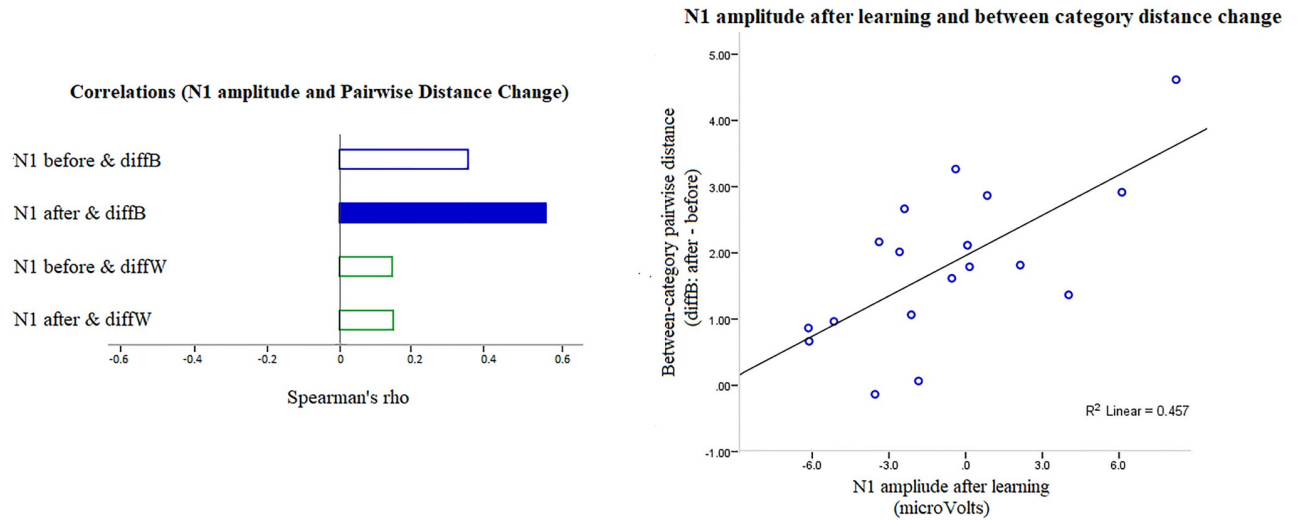

**Fig 10. The relationship between N1 and CP separation in Learners. Left:** Spearman rank order correlations between N1 amplitude before and after learning and changes in pairwise distance between (diffB) and within (diffW) categories. The only significant correlation is the dark bar. **Right:** Scatter plot for the significant correlation: Decreased N1 amplitude (peaks that were less negative) predicted an increase in between-category distance.

https://doi.org/10.1371/journal.pone.0226000.g010

subjects who had succeeded in reaching our learning criterion (80%) (Learners). There were no effects with the Non-Learners. These effects were replicated in Experiment 2: Categorization training induced between-category separation and significant "Global CP" in the Learners and no effect in the Non-Learners. N1 amplitude also correlated with the size of the perceived between-category separation (a CP effect) after learning (again in Learners only) whereas LPC amplitude correlated with categorization accuracy and Reaction Time. To our knowledge, this study provides the first report of a correlation between an electrophysiological index of category learning and a behavioral index of categorical perception.

### Perceptual effects of category learning

To learn to categorize requires detecting the features that distinguish the members of the category from the non-members: the features that covary with category membership/non-membership. Experiment 2 replicated a perceptual change that had occurred only in those subjects who had successfully learned to distinguish the members of two different categories after 45 minutes of trial and error training with corrective feedback; the change was absent in those who had had exactly the same training trials but had failed to learn to distinguish the categories. This perceptual change, categorical perception (CP), consisted of an increase in dissimilarity between the members of different categories ("between-category separation") after successfully learning to categorize them, sometimes also accompanied by a weaker decrease in dissimilarity between members of the same category ("within-category compression").

The direct function of categorization is to differentiate members from non-members, so as to be able to do the right thing with the right kind (category) of thing. This requires selectively detecting the features that distinguish the members from the non-members and ignoring the features that do not distinguish them. Hence increased between-category distinctiveness (separation) would be a direct effect of category learning whereas increased within-category similarity (compression) would only be a side-effect. This may be the reason why between-category separation is stronger than within-category separation in learned CP.

## The role of unsupervised learning

Our results also show that unsupervised or unreinforced learning (i.e., mere passive exposure, without feedback as to what to do with what) can also play role in visual category learning. In Experiment 2, the subjects who would eventually turn out to be successful Learners after training were already rating between-category pairs as significantly more dissimilar than within-category pairs before any supervised learning (trial and error with corrective feedback) had begun, hence before they had learned the categories. (Six of them proved to be "immediate Learners," already above our 80% criterion as soon as the training trials began). Furthermore, the N1 amplitudes before reaching learning criterion were already different from the N1 amplitudes of "Non-Learners" This suggests that the passive exposure during the dissimilarity judgments themselves was already inducing some selective feature detection based on feature frequencies and intercorrelations intrinsic to the stimulus space itself.

## Dimensional reduction and neural network model

The selective detection of the features that distinguish the members of a category from the non-members can be understood as a form of *dimensional reduction* [19,24,58,59]: Stimulus features can be treated as dimensions of a (discrete) N-dimensional similarity space. If we compare two stimuli, our initial perception of similarity is based on all N of these dimensions. Category learning occurs when the subject has successfully detected the k features that covary with category membership, ignoring the N-k non-covariant features as irrelevant. This reduced k-dimensional subspace of the original N-dimensional similarity space acts as a feature filter, hence a potential mechanism underlying the changes in the inter-stimulus distances after learning a category.

This potential explanation for the observed learned CP effect was tested in a neural network model sketched in a previous paper [48] and is being further developed in a forthcoming work in preparation [57]. The net generates CP through two learning stages, the first unsupervised (autoencoding of binary N-dimensional stimuli) and the second supervised (sorting the stimuli into two categories through trial and error with corrective feedback). The net has a layer of h "hidden units" between input and output whose activation values can vary from -1 to 1. The distances between and within the categories before and after category learning can then be calculated as the average euclidean distance between the hidden-unit representations. The CP effect is then computed by subtracting the average distances between and within categories before and after the categories have been learned. Dimensional reduction is confirmed by tracking the weights accorded to dimensions by the network in feature space (hidden unit-activation space) before and after learning.

This preliminary neural model indicates that CP can be modeled as the difference between the outputs of perceptual filters (i.e. feature-detectors) learned through (1) unsupervised learning and then fine-tuned through (2) supervised learning. There is learning in both phases, but the net starts out as a randomly connected blank slate. Unsupervised learning (mere passive exposure to samples from the infinite space of possible stimuli) allows it to learn the "lay of the land" from the intrinsic frequencies and correlations among the stimulus features, but without any feedback about categorization (Fig 11). Categorization is learned in the supervised phase. With human subjects it is assumed that they have already had a lifetime of passive exposure to stimuli and that the textures in this experiment, though novel, will nevertheless not be completely unfamiliar stimuli. During the dissimilarity judgments before categorization training the subjects are exposed to the stimuli, and some learning based on feature frequencies and correlations probably takes place too.

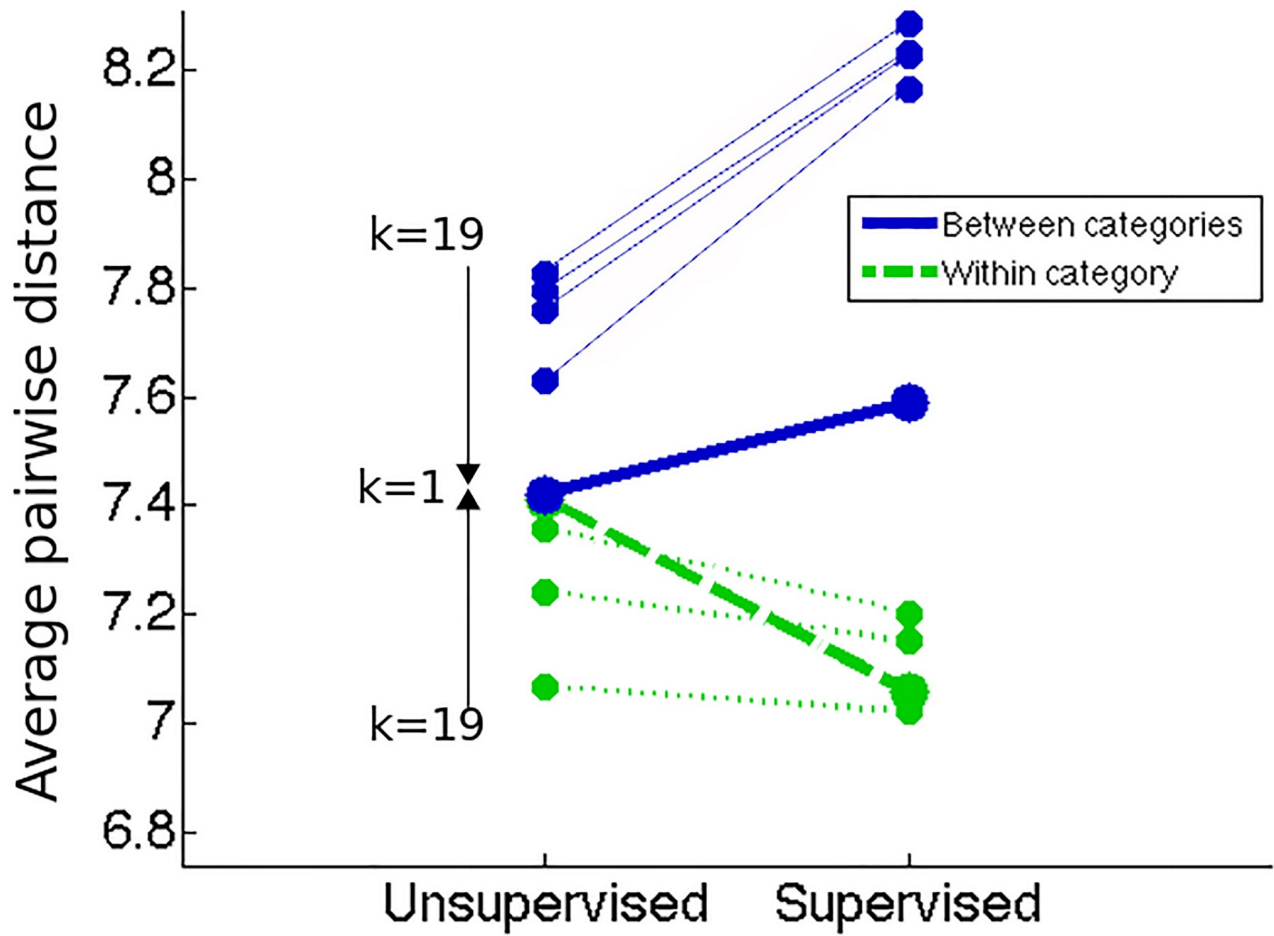

**Fig 11. Average pairwise distance between and within categories measured first before supervised learning by the neural nets and then after for stimuli with different proportions of covariant features (k).** The learning by the nets generated a robust pattern of increasing between-category distance from before supervised learning to after (separation, blue lines). This was found for the Learners in all the human experiments. The net also consistently generated decreasing within-category distance (within, green lines). This was found in some of the human Learners. There was another clear pattern in the nets that was present only very approximately in our human experimental subjects: For stimuli with a higher proportion of k, the unsupervised learning (autoencoding) before the supervised learning is already enough to generate some initial differences similar to those observed in human subjects (cf. Fig 5).

https://doi.org/10.1371/journal.pone.0226000.g011

The model also shows that near perfect categorization rates can be reached when the networks are given unlimited training. However, for a fixed amount of training, different levels of categorization performance can be observed among networks. To assess whether the human CP effect can be accounted for by this model, a threshold for successful learning was set to a regularized mean squared error of $10^{-3}$ on the last layer of the network and nets were labeled as Learner and Non-Learner nets. The model predicts (and potentially explains) two of our three principal experimental effects: (a) the human CP effects of between-category separation (and sometimes also within-category compression) in successful Learners (Fig 11), and (b) the correlation of the CP with categorization accuracy (bigger CP for better performance).

In the nets, reducing the proportion of covariant features (k) makes the task more *difficult or complex*, in the sense that there is more irrelevant variation in the data: with all other parameters equal, the network requires more computation to categorize successfully (100% success). The net is hence a potential model for the mechanism underlying category learning. It also supports the hypothesis that dimensional reduction is the mechanism underlying CP: After the supervised learning phase, the hidden unit activations for the N dimensions are transformed, weighting the covarying dimensions more heavily (reduction from N to k).

However, unlike the nets, for which visual stimuli are just N-component vectors of binary features, the human visual system does not process all features equally: some are more salient or detectable than others [48]. Testing the dimensional reduction hypothesis in human subjects would thus require either (a) an extremely large number of experimental conditions and subjects to try to counterbalance for feature inequalities or (b) an attempt to simplify the features to make them more equal. The neural net provides a means of running many simulations of different conditions in short periods of time, helping to model the explanations and effects seen in human subjects. We hope it will help analyze the relationship between the number of covarying features (k/N) and the size of the resulting CP effect under different conditions without the bias of feature inequalities.

### ERP changes

Our analysis showed that two ERP components change significantly from before to after reaching the 80% criterion: the Late Positive Component (LPC), which is presumed to be decisional, and the first negative occipital peak (visual N1), which is presumed to be perceptual. The late parietal LPC peaks between 600 and 800 ms after a stimulus. It is thought to reflect higher-order cognitive processing: conscious recollection [60,61], memory-related judgments [55,62], and retrieval success [63,64]. It has also been shown to be sensitive to decision accuracy [65] and response confidence [66].

In both experiments, we observed a significant increase in LPC positivity from before to after training only in those subjects who successfully reached our 80% criterion (i.e., the Learners). Given its time course and existing functional interpretation, these LPC changes accompanying category learning cannot account for our perceptual effects. The LPC increase correlated with measures of successful learning: reaction times and decision accuracy. This is consistent with prior findings of a larger LPC for correct responding in a category learning task [47]. Overall, we interpret the increase in the LPC positivity as a correlate of later-stage cognitive processing affected by category learning, including the link between perceptual information stored in working memory and potential responses as well as the selection, execution and conscious recollection of a strategy.

The visual N1 is defined as the first negative occipito-temporal component evoked by any complex visual stimulus, peaking between 150 and 200 ms. It is thought to reflect visual discrimination [67], feature selection [68], changes in visual attention [54,69]; and facilitation of the processing of task-relevant information [70]. In both experiments, we observed N1 negativity decreases throughout the categorization task in Learners only. This suggests that learning a category induces changes in early perceptual processes such as feature extraction or selective attention. The possibility that this N1 decrease is due to repeated exposure to the stimuli is ruled out by the fact that the only difference between the Learners and the Non-Learners was that the Learners had successfully learned to categorize (in the 45-minute training session) whereas the Non-Learners had not. We infer that the N1 negativity decrease is related somehow to the feature filtering that underlies categorical perception, indexing either feature extraction or changes in attentional weighting that highlight the features that covary with category membership.

Another interesting finding regarding N1 is that whereas the pre-learning N1 values did not differ between Learners and Non-Learners in Experiment 1 (in which only categorization training was done), they differed significantly in Experiment 2, in which subjects did pairwise dissimilarity ratings both before and after training. These differences in Experiment 2 suggest, once again, that before categorization training some of the (eventual) Learners were already processing the stimuli differently from the Non-Learners as a side-effect of the passive exposure (unsupervised learning) in making their pairwise dissimilarity judgments. This exposure could already have induced a change in the way subjects processed the features, making the relevant features stand out more for these early and immediate Learners.

Category-related effects occurring between 100 and 150 ms have already been reported in the N1 literature in relation to categorization: more N1 negativity for (a) category members than for non-members after learning a category [71]; (b) for natural categories than artifactual ones [72]; (c) for subordinate-level categories than basic levels [44]; (d) for correct than incorrect trials in category learning based on information integration—and the opposite (more negativity for correct than incorrect trials) for rule-based category learning [47].

The studies reporting N1 effects did not test directly for learned CP Effects, however. In the series of studies reported here, the size of the CP (between-category separation) increased as N1 negativity decreased. Because the N1 is a negative deflection, this pattern corresponds to an amplitude reduction: Less negative N1 amplitudes after learning predicted increased CP. This is a correlation between an objective, physiological measure resulting from category learning—a perceptual ERP component—and a subjective measure of perceptual dissimilarity (CP separation). It shows that learning a category can induce changes in perceptual processing even after only a 45-minute training period. Future studies will need to examine the effects of long-term distributed training and overlearning (see below).

In an fMRI study of category learning, Folstein reported that learning a category enhanced the discriminability of category-distinguishing features in the extrastriate visual cortex of human subjects [43]. This finding was accompanied by increased discriminability of relevant dimensions measured after training. This agrees with our observed changes in the N1 component, which source localization has traced to the extrastriate cortex, in the ventral visual pathway [68,73]. This provides further support for the perceptual basis of CP.

Folstein and others have shown that CP is an effect of "dimensional separation" through attentional weighting of relevant features/dimensions that is not specific to category category boundaries, rather than being a category boundary effect [74]. Dimension reduction and category boundary effects are not contradictory. Differential attentional weighting of distinguishing (category-covariant) versus non-distinguishing features or dimensions is an important component of category learning and dimensional reduction, as exemplified by our neural network model.

The case of color perception—in which categories are compression bands and boundaries along a single wave-length dimension—is not representative of most learned categories, where it is entire dimensions (features) that are either attended or ignored [33]. Between-category boundaries do not occur along a single stimulus dimension in most category learning; rather, they are separation/compression effects in multidimensional subjective similarity space, which is reduced from the N total dimensions before learning to k category-covariant dimensions after learning. Attempts to generate multiple CP boundaries along a single stimulus dimension (as opposed to just the midpoint, [75] through learning in human subjects have had limited success [12,76].

### ERPs, name bias and the subjectivity of dissimilarity judgments

CP effects have been criticized as being artifacts of verbal bias [37,77–79]. The suggestion has been that the members of different categories do not really look more different after we learn to categorize them (nor do members of the same category look more similar); the difference in dissimilarity ratings merely reflects the fact that members of different categories are associated with different names and members of the same category are associated with the same name: it is just the sameness or difference of category names that biases our judgment as to how similar they are. What this criticism overlooks is the fact that being able to associate a correct category name with a member of a category is not just a matter of pinning a name on it, as in the paired-associate learning of nonsense syllables pairs, or the pairing of nonsense syllables with individual pictures. With a large or infinite variety of inter-confusable members and non-members successful categorization can require laboriously learning to distinguish the members from the nonmembers by trial and error, as our (successful) subjects had to do, by discovering the features that distinguish them.

It hence seems more likely that it is the process of learning to distinguish the categories that influences their similarity, rather than just the name that is associated with the members once a Learner has succeeded in learning to categorize them [80]. Moreover, in our studies what was associated with each category was just a key-press (K or L) rather than overt naming. Parallel studies in our laboratory have also found that the same pattern of separation/compression occurs whether one uses subjective dissimilarity ratings or objective psychophysical discriminability measures (ABX) and signal detectability analysis (d') to measure interstimulus distance [81]. The fact that we observed a positive correlation between the size of the separation effect and the size of the N1 component after learning but no correlation with the later, verbal/decisional LPC further reinforces our conclusion that our CP effects are a reflection of the perceived distinctness of the members of the two categories, rather than just a bias from the distinctness of their names.

### Early vs. late learning

Our arbitrary learning criterion was 16/20 (80%) correct across a block of 20 consecutive trials and remaining above that 80% rate for all subsequent blocks (with at least two more 20-trial blocks remaining till the end of training). It is likely, however, that the learning in this single 45-minute training session of 400 trials is a matter of degree rather than all-or-none, as the correlational results of our alternative continuous analysis show (Fig 6). Moreover, the number of trials before and after learning was unequal in almost all cases. Some subjects learned early in the 400-trial series, some learned late. Our averages across subjects included a subject who learned after 80 trials (yielding an 80/320 split) and another who learned only after 280 trials (280/120 split). This affected our variance (making it bigger for those conditions with a smaller number of trials) and added noise to our averaged ERP data because the Signal to Noise Ratio was smaller for ERPs based on a smaller number of trials. To address this issue, we did a second, supplementary analysis, summarized in S1 File, in which we analyzed the changes across the four blocks of 100 trials. The same N1 and LPC effects emerged throughout the blocks. In testing for statistical significance, the main significant change emerged in comparing the first to the last block. Most subjects had not learned in the first block but were already categorizing successfully in the last one. This outcome confirms our before-after analysis.

### Conclusions

Our results provide both behavioral and electrophysiological evidence that learning a new category can induce changes in perception. Increases in the perceived difference between categories (between-category separation) after successful learning were more prominent than within-category compression. This makes sense because detecting the differences between members of different categories is more important than ignoring the differences within the same category. Our ERP results suggest that learning a category can influence early stages of visual perception, as indicated by decreased N1 negativity after successful learning. The fact that the size of this perceptual component after learning the category was positively correlated with the between-category separation (reduced N1 negativity predicting greater perceptual separation) further reinforces our evidence that category learning can modify perception. If learning the meaning of words from words, indirectly [82], is grounded in direct sensorimotor learning to categorize their referents in the world, then our findings are also evidence of a subtle Whorfian effect of language on how we see the world.

## Supporting information

**S1 File. Appendix.** Supplementary analysis by block of Experiments 1 and 2.
(DOCX)

**S1 Fig. ERP waveforms and scalp maps for the four successive 100 trial blocks in Experiment 1 (occipital electrodes and N1 window).**
(TIF)

**S2 Fig. ERP waveforms and scalp maps for the four successive 100 trial blocks in Experiment 1 (parietal electrodes and LPC window).**
(TIF)

**S3 Fig. ERP waveforms and scalp maps for the four successive 100 trial blocks in Experiment 2 (occipital electrodes and N1 window).**
(TIF)

**S4 Fig. ERP waveforms and scalp maps for the four successive 100 trial blocks in Experiment 2 (parietal electrodes and LPC window).**
(TIF)

**S1 Data. Database for Experiment 1.**
(XLSX)

**S2 Data. Database for Experiment 2.**
(XLSX)

## Author Contributions

**Conceptualization:** Fernanda Pérez-Gay Juárez, Stevan Harnad.

**Data curation:** Fernanda Pérez-Gay Juárez, Tomy Sicotte.

**Formal analysis:** Fernanda Pérez-Gay Juárez, Tomy Sicotte, Christian Thériault.

**Funding acquisition:** Stevan Harnad.

**Investigation:** Fernanda Pérez-Gay Juárez, Tomy Sicotte, Christian Thériault.

**Methodology:** Fernanda Pérez-Gay Juárez, Tomy Sicotte, Christian Thériault.

**Project administration:** Fernanda Pérez-Gay Juárez.

**Software:** Tomy Sicotte, Christian Thériault.

**Supervision:** Stevan Harnad.

**Validation:** Fernanda Pérez-Gay Juárez, Stevan Harnad.

**Visualization:** Fernanda Pérez-Gay Juárez.

**Writing – original draft:** Fernanda Pérez-Gay Juárez.

**Writing – review & editing:** Fernanda Pérez-Gay Juárez, Stevan Harnad.